\documentclass[runningheads]{llncs}
\usepackage[T1]{fontenc}
\usepackage{comment}
\usepackage{graphicx}
\usepackage{subcaption}
\usepackage{changepage} 
\usepackage{rotating}
\usepackage{multirow}
\usepackage{xcolor}
\usepackage{booktabs}
\usepackage{amsmath}
\usepackage{enumitem}

\begin{document}
\title{Do we need more complex representations for structure? A comparison of note duration representation for Music Transformers}
\titlerunning{Do we need more complex representations for structure?}
%
\author{Gabriel Souza\inst{1} \and
Flavio Figueiredo\inst{1}\and
Alexei Machado\inst{2} \and
Deborah Guimarães\inst{1}
}
\authorrunning{G. Souza et al.}
\institute{$^1$Department of Computer Science, Universidade Federal de Minas Gerais  \\
$^2$Department of Computer Science, Pontifical Catholic University of Minas Gerais\\Belo Horizonte MG, Brazil,
\url{https://dcc.ufmg.br/}}
\maketitle              
\begin{abstract}
In recent years, deep learning has achieved formidable results in creative computing. When it comes to music, one viable model for music generation are Transformer based models. However, while transformers models are popular for music generation, they often rely on annotated structural information. In this work, we inquire if the off-the-shelf Music Transformer models perform just as well on structural similarity metrics using only unannotated MIDI information. We show that a slight tweak to the most common representation yields small but significant improvements. We also advocate that searching for better unannotated musical representations is more cost-effective than producing large amounts of curated and annotated data.\\
\textbf{keywords:} Music Composition, Creative Computing, Deep Learning
\end{abstract}
\section{Introduction}
From simple Markovian models~\cite{Brooks1957} to the most recent Deep Learning Transformer architectures~\cite{Huang2018,Wu2020,Shih2022}, automatic music generation systems have gained significant performance. Depending on the setting, automatically generated music is able to fool people into believing that such scores were made by humans~\cite{Huang2018}.
Nevertheless, even state-of-the-art models struggle to generate convincing musical pieces with longer duration due to its self-referential structure. A musical phrase lasting a few seconds may contain variations of a few melodic elements, whereas a musical section lasting a minute may contain variations of few phrases. The piece itself may contain variations of a few sections.

Over the last few years, authors exploring transformer architectures have attempted to incorporate structural information into their models using either external annotations~\cite{Wu2020} (i.e. annotations of music segments or bars), or changes in the model's architecture~\cite{Shih2022} in order to incorporate similarity.

\begin{figure}
\includegraphics[width=\textwidth]{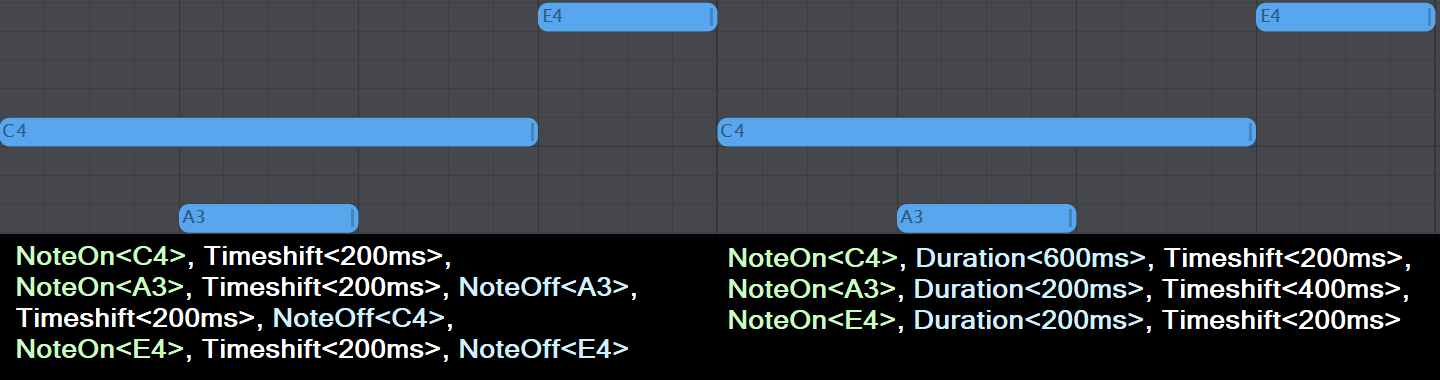}
\caption{An example of the same stretch of music notated on the traditional way (left) and with explicit duration (right).} \label{f:ex}
\end{figure}

In this paper, we inquire if the effort of annotating musical data is worth the trouble by showing that a simple change to representation style, without adding new information, yields similar or better improvements when compared to using annotated data. Figure~\ref{f:ex} illustrates the difference between these unannotated representations. Instead of using \texttt{NoteOn} and \texttt{NoteOff} events, we train the model with \texttt{NoteOn} and \texttt{Duration} events, similar to one of the styles found in Fradet's work \cite{ismir2023}. This approach avoids placing related \texttt{NoteOn} and \texttt{NoteOff} events too far apart.

This representation is promising because it applies to any MIDI dataset. Moreover, through a statistical analysis of four different datasets, we show that structural similarity measures (discussed in Section~\ref{sec:methods}) are overall improved with this approach, with little to no impact on other metrics. Our results are backed by a user study where listeners are generally more able to identify coherent musical structure based on explicit notation.
In short, our argument is that external annotations may not be worth including in transformer models. This result is particularly useful when creating large musical datasets and attempting to generate longer pieces with a coherent structure.

\section{Related Work}
Researchers have early realized that the choice of musical data representation was decisive in model performance. In a pioneer work, Todd~\cite{Tod1989} explored recurrent neural networks (RNNs) and generated monophonic melodies. Chen and Miikkulainen~\cite{Che01} also produced short musical pieces using deep learning with limited results.


Regarding musical representation, one-dimensional note threads have been the canonical choice for music representation in Hidden Markov Models and Long Short Term Memory networks (LSTM)~\cite{Oor18}, with polyphonic music requiring the serialization of the multiple tracks into a single sequence, as adopted by Dong {\em et al.}~\cite{Don18}. 
More recently, Jiang and coworkers~\cite{Jia22} have proposed a hierarchical model composed of multiple encoders that attempt to learn a latent representation of the bars, while attention blocks capture the global structure of the music. Multi-track representation was further developed by Shih {\em et al.} in the Theme Transformer~\cite{Shih2022}, and by Dong {\em et al.}~\cite{Don23} pushing the limits of representation towards the generation of orchestral music.

The quest for longer sequences has been followed by the urgency for a more quantitative evaluation of musical quality. In their work on Jazz music generation~\cite{Wu2020}, Wu and Yang propose objective metrics to assess structuring, pitch usage, rhythm and harmony. 

\section{Methods and Materials}\label{sec:methods}
Datasets containing varied musical genres, construction methods and musical complexities were selected for this work, with the goal of verifying if explicit representation offers improvement on all kinds of data. Every piece had their percussive elements removed and all instruments were combined into a single piano track.
The four selected datasets are:
(a) the MAESTRO Dataset\footnote{\url{https://magenta.tensorflow.org/datasets/maestro}}, containing a collection of classical piano pieces performed by contestants of the International Piano-e-Competition, transcribed into midi files by the electronic piano used in the competition;    (b) the VGMusic-SNES Dataset\footnote{\url{http://vgmusic.co/music/ console/ nintendo/snes/}}, a collection of soundtracks from games for the old Super Nintendo video-game console; 
    (c) the Weimar Jazz Database\footnote{\url{https://jazzomat.hfm-weimar.de/dbformat/dboverview.html}}, which contains transcriptions of jazz solos; and (d) POP909\footnote{\url{https://github.com/music-x-lab/POP909-Dataset}}, a dataset containing professional piano arrangements of popular songs.
 
\begin{table}[t!]
    \centering
    \caption{Some characteristics of the datasets on the original (left) and explicit (right) notation, including average length of each piece in tokens, and average number of unique tokens in each piece.}
    \begin{minipage}{0.48\textwidth}
        \centering\small
        \begin{tabular}{lrrrr} \toprule
        & \textbf{Maestro} & \textbf{SNES} & \textbf{Jazz} & \textbf{Pop} \\ \midrule
        Total Files & 1282 & 6591 & 455 & 909 \\
        Train Files & 967 & 5937 & 413 & 849 \\
        Val. Files & 137 & 501 & 29 & 36 \\
        Test Files & 178 & 154 & 13 & 24 \\
        Avg. Length & 22820 & 7316 & 2147 & 6831 \\
        Avg. Uniques & 231 & 95 & 125 & 149 \\
        \bottomrule
        \end{tabular}

        \label{t:data}
    \end{minipage}\hfill
    \begin{minipage}{0.48\textwidth}
        \centering\small
        \begin{tabular}{lrrrr} \toprule
        & \textbf{Maestro} & \textbf{SNES} & \textbf{Jazz} & \textbf{Pop} \\ \midrule
        Total Files & 1282 & 6591 & 455 & 909 \\
        Train Files & 967 & 5937 & 413 & 849 \\
        Val. Files & 137 & 501 & 29 & 36 \\
        Test Files & 178 & 154 & 13 & 24 \\
        Avg. Length & 22608 & 5641 & 1692 & 6119 \\
        Avg. Uniques & 267 & 70 & 140 & 186 \\
        \bottomrule
        \end{tabular}
        \label{t:data-duration}
    \end{minipage}
\end{table}

Regarding metrics, objective evaluation of musical quality is currently an open problem with no standardized benchmarks. Even more so when evaluating musical pieces of different genres, broadening the scope of the challenge and excluding the possibility of measuring similarity to a pre-defined set of musical conventions found in a certain genre or followed by a specific composer. Considering the intrinsically subjective experience of music, the problem may very well be unsolvable.

Still, Western tonal music has some general conventions and patterns that \emph{can} be measured and used as a reasonable proxy for musical quality. Works like~\cite{Wu2020,Yang2018a,Ens2018,Ens2020c} propose a few ways to capture the complexity of musical structure in measurable metrics.

Four structural metrics were used to evaluate the models: Structureness Indicators and Pitch Class Entropy, also explored by the Jazz Transformer~\cite{Wu2020}; Pitch Class Consistency, explored by the Theme Transformer~\cite{Shih2022}; and Compression Ratio, as described by Chuan and Herremans~\cite{Chuan_Herremans_2018}. As for the subjective evaluation, we decided to collect the same measurements proposed on the Jazz Transformer~\cite{Wu2020} on Likert-5 Scale, namely: Overall Quality (O), Impression (I), Structureness (S), Richness (R). To that end, the following form, mostly identical to the one from \cite{Wu2020}, was sent to voluntaries: \\

{\it ``In this form, you will answer questions regarding a musical piece (song). The music that you will hear has a short initial part composed by human beings. The exact size of the human-composed snippet is not disclosed to participants, so please do not assume any value. Just know it is a few seconds. After this initial excerpt, the rest of the song (a few minutes) was generated by AI.\\
            \\
Musical pieces are presented in a simple format called MIDI. So for your answers, please try not to make a comparison with professionally recorded music. Knowing this information, please let us know if you identify themes that are repeated in the song, in addition to your general opinion of it, according to the following questions (answers to all questions are mandatory):

\begin{enumerate}[label=(\alph*)]
\item \textbf{Overall Quality (O):} Does it sound good overall?
\item \textbf{Impression (I):}      Can you remember a certain part or the melody? That is, is the major theme of the initial seconds present in the overall piece?
\item \textbf{Structureness (S):}   Does it involve recurring music ideas, clear phrases, and coherent sections? This recurrence may occur in any part of the piece.
\item \textbf{Richness (R):}        Is the music diverse and interesting?
\end{enumerate}}

\section{Experimental Results}
The experiments consisted in using the Music Transformer~\cite{Huang2018} as a baseline, and substituting their music representation with the explicit notation exemplified in Fig.~\ref{f:ex}. A PyTorch implementation of the model\footnote{\url{https://github.com/gwinndr/MusicTransformer-Pytorch}} was trained with the recommended hyper-parameters (see~\cite{Huang2018}). Models with both notations were trained for 100 epochs for the Maestro dataset. For every other dataset, the models were trained for 20 epochs, since they contain less data or simpler data. In total there were eight models, one for each dataset and input representation.

After training, 30 pieces were generated per dataset\footnote{\url{https://drive.google.com/file/d/1qZp4wAxqcgE2w-u4h9xZnKw1TY3cyIY-/view?usp=sharing}} 
by randomly choosing ten files from the test dataset, giving to the model the first 256 tokens of those files as primers, and letting the model generate three continuations for each one by completing the sequence until it had 2048 tokens. 

The subjective evaluation form was answered by 123 participants. Evaluation was stopped when at least 100 forms were obtained for each notation. Considering a tuple of notation and dataset, each pair was evaluated by at least 20 participants.

\subsection{Structural Metrics}
The results for the Structureness Indicators (SI) are shown in Table~\ref{t:si}. The SI are extracted from the fitness scape plot~\cite{Muller2015}. This plot extracts the degree of repeated structures on a musical piece and is extracted from the self-similarity matrix (SSM). The measure ranges from 0 to 1, with higher values indicating more repetition.
Moreover, as proposed by the original authors, SI is measured for short-, medium-, and long-term duration. In this sense, we used 3 to 8-second intervals (short-), 8 to 15 intervals (medium-), and 15-second intervals (long-). In order to obtain more representative results, given the randomness involved in creative computing using deep learning, we bootstrap~\footnote{Bootstrap CIs were chosen as these metrics do not have statistical tests related to them.}\cite{Wasserman2004} from the generated pieces by re-sampling 9999 times and using the bias-corrected and accelerated method by \cite{71c4d02a-0db0-31e8-95b7-4bdc20b71cf9}.

Initially, we point out that the explicit notation shows improvements overall. A higher impact is present when measuring SI on the long-term. Nevertheless, statistical significance is only present on the Jazz and Pop datasets (where improvements range from 10\% to 97\%). These results are interesting for two reasons. Firstly, the overall improvements are higher than those reported by~\cite{Wu2020} (we used the same code) for Jazz. This indicates that explicit notation is actually better than using complex notations for SI. More notably, our results increase as the piece becomes longer.

Next, we turn our attention to the Pitch Class Entropy. This metric is simply Shannon's entropy, defined as Equation~\ref{eq:entropy} when measured on the 12-dimensional pitch class histogram.
\begin{equation}
    \label{eq:entropy}
    H(p) = -\sum^{12}_{i=1} p(i) \log_2(p(i))
\end{equation}
Table~\ref{tab:combined} reveals that there is only one {\em small} statistical difference (Pop) comparing the original and explicit notations. This indicates that both approaches perform just as well, showing that the explicit notation will not reduce this score.

Following that, we present the Pitch Class Consistency results, shown in Table~\ref{tab:combined}.This metric helps in determining the consistency of tonality along the piece, helping with some of the ambiguity inherent to the pitch class entropy. It is computed via the Kullback-Leibler divergence between consecutive Pitch Class histograms when a musical piece is split in equal-sized windows.
Similar to the results for Entropy, overall notations perform equally, with one exception occurring for Jazz pieces. Overall, explicit notation presents an increase in Pitch Class consistency similar to the one observed by the Theme Transformer~\cite{Shih2022} (though the authors are not clear on which divergence metric they use).

Finally, Table~\ref{tab:compression_ratio} shows a similar result to the Pitch Class Entropy results: using the explicit notation results in significantly higher compression ratios for the pop dataset, but no significant changes on any other.

\begin{table*}[t]
\caption{Structureness Indicators for the original and explicit notation models. Values expressed as 95\% C.I.}
\centering\small
\resizebox{\textwidth}{!}{
    \begin{tabular}{lcccccccccccc}
    \hline
    & \multicolumn{3}{c}{Original} & \multicolumn{3}{c}{Explicit} & \multicolumn{3}{c}{Improvement} \\
    Dataset& Short & Medium & Long & Short & Medium & Long & Short & Medium & Long \\ \hline
    Maestro & (0.32, 0.37) & (0.18, 0.23) & (0.07, 0.13) & (0.31, 0.37) & (0.19, 0.27) & (0.10, 0.19) & -0.56\% & 9.08\% & 29.69\% \\
    SNES & (0.30 0.39) & (0.14, 0.25) & (0.08, 0.17) & (0.37, 0.45) & (0.25, 0.36) & (0.16, 0.26) & 17.44\% & 61.93\% & 71.56\% \\
    Jazz & (0.27, 0.29) & (0.23, 0.24) & (0.16, 0.19) & (0.28, 0.31) & \textbf{(0.25, 0.29)} & \textbf{(0.20, 0.24)} & 3.54\% & \textbf{10.49\%} & \textbf{22.92}\% \\
    Pop & (0.29, 0.32) & (0.21, 0.25) & (0.11, 0.16) & \textbf{(0.34, 0.42)} & \textbf{(0.28, 0.37)} & \textbf{(0.23, 0.31)} & \textbf{23.85\%} & \textbf{40.83\%} & \textbf{97.86\%} \\ \hline
    \end{tabular}
}
\label{t:si}
\end{table*}

\begin{table}[tb]
\centering
\caption{Pitch Class Entropy and Consistency of the music generated by the original and explicit notation models. Values expressed as 95\% C.I.}
\begin{tabular}{llcccc}
\hline
& & \multicolumn{2}{c}{Original} & \multicolumn{2}{c}{Explicit} \\
& Dataset & Real & Generated & Real & Generated \\
\hline
\multirow{4}{*}{\begin{sideways}{\scriptsize Entropy}\end{sideways}} & Maestro & (3.00, 3.35) & (3.04, 3.17) & (2.99, 3.34) & (3.09, 3.28) \\
& SNESS & (2.36, 2.91) & (1.55, 2.03) & (2.38, 2.92) & (1.96, 2.34) \\
& Jazz & (3.18, 3.38) & (3.35, 3.41) & (3.18, 3.38) & (3.41, 3.47) \\
& Pop & (2.76, 2.91) & \textbf{(2.65, 2.73)} & (2.76, 2.91) & \textbf{(2.87, 2.97)} \\
            \hline
\multirow{4}{*}{\begin{sideways}{\scriptsize Consistency}\end{sideways}} & Maestro & (0.33, 1.07) & (2.10, 2.93) & (0.31, 1.19) & (1.79, 3.28) \\
& SNES & (0.48, 1.35) & (1.13, 2.31) & (0.47, 1.24) & (0.91, 1.64) \\
& Jazz & (1.22, 3.00) & \textbf{(0.77, 0.99)} & (1.23, 3.05) & \textbf{(0.47, 0.67)} \\
& Pop & (0.16, 0.37) & (0.91, 1.32) & (0.16, 0.38) & (0.77, 1.02) \\
            \hline
\end{tabular}
\label{tab:combined}
\end{table}

\begin{table}[t!]
    \centering
    \caption{Compression Ratios for pieces generated using the original and explicit notation models. Values expressed as 95\% C.I.}
    \begin{tabular}{lcc}
        \hline
        Dataset & Original & Explicit \\ 
        \hline
        Jazz & (1.39, 1.41) & (1.41, 1.43) \\ 
        Maestro & (1.42, 1.45) & (1.36, 1.42) \\ 
        POP909 & (1.44, 1.50) & \textbf{(1.58, 1.66)} \\ 
        SNES & (2.62, 3.07) & (2.79, 3.43) \\ 
    \end{tabular}
    \label{tab:compression_ratio}
\end{table}

\subsection{Listening Evaluation}
In Table~\ref{tab:user2} we can see that the explicit notation has five statistical wins against two for the original notation. When considering structure (S), results are mostly achieved on SNES and Pop, more repetitive datasets by their nature. Nevertheless, it is interesting to see that under the explicit notation, Maestro and Pop pieces are overall of more quality (O). Finally, Jazz notations cause an overall larger impression (I) under the explicit notation. These results support our hypothesis that tangible improvements can be observed with simple changes in input representation.

The last line contains the average result for each question aggregated by dataset. Statistical significance of results, was tested with a MannWhitney-U test and is shown in bold. Here, we can initially see that overall scores are low. This is somewhat expected as we did not cherry-pick any musical piece for the study. Nevertheless, listeners are able to identify more structure on the explicit notation (providing evidence towards our hypothesis).

\begin{table}[t]\centering
    \caption{Average results in Likert-5 Scale for the listening study per dataset and notation, as well as overall average. Statistically significant values when comparing Explicit and Original notations are shown in bold ($p<0.05$).}
    \centering
    \begin{tabular}{lcccccccc}
    \hline
        & \multicolumn{4}{c}{Original} & \multicolumn{4}{c}{Explicit} \\
        Dataset & O & I & S & R & O & I & S & R \\
        \hline
        Maestro & 3.26 & 3.19 & 3.55 & 3.16 & \textbf{3.79} & 3.38 & 3.38 & 3.71 \\
        SNES    & \textbf{1.86} & \textbf{2.71} & 3.39 & 1.79 & 1.62 & 2.26 & \textbf{3.85} & 1.44 \\
        Jazz    & 2.50 & 2.71 & 3.21 & 2.58 & 2.42 & \textbf{3.31} & 3.23 & 2.54 \\
        Pop     & 2.55 & 2.64 & 2.95 & 2.41 & \textbf{3.35} & 3.20 & \textbf{3.75} & \textbf{3.20} \\
        Average & 2.56 & 2.84 & 3.30 & 2.50 & 2.61 & 2.93 & \textbf{3.58} & 2.52 \\
        \hline
    \end{tabular}
    \label{tab:user2}
\end{table}

\section{Conclusion}
In this work, we demonstrated how a simple change to input notation impacts the structural quality of AI-generated music, sometimes more than adding information to the representation. Considering the time and computational costs associated with transformers and other large architectures - with memory usage and processing time proportional to sequence length and number of token types - any improvements that come from simple input formatting deserve study.

Even when costs are not a concern, the recent success of LLMs that train on massive amounts of data without semantic annotations is aligned with our findings, suggesting that these annotations may not be necessary in this domain either.

As such, future works on music generation using transformers should go in the same direction as text generation: massive datasets that don't include any human annotation, and thus are cheaper to obtain. In addition, other tokenization techniques could be explored. Polyphony and note duration quantization are potential targets for improvements as these are usually responsible for most of the complexity when tokenizing music.

\section*{Acknowledgements} The authors acknowledge funding from CAPES through travel support to present this paper. We also acknowledge FAPEMIG for a scientific initiation scholarship.

%
%
%
\bibliographystyle{splncs04}
\bibliography{bibs}

\end{document}